\documentclass[a4paper,12pt]{article}

\usepackage{graphicx}
\usepackage{grffile}
\usepackage{amssymb}
\usepackage{amsmath}

\usepackage[usenames,dvipsnames]{color}
\usepackage{verbatim}
\usepackage{hyperref}
\usepackage{setspace}

\newcommand\T{\rule{0pt}{2.2ex}}
\newcommand\B{\rule[-1.0ex]{0pt}{0pt}}

\newcommand{\BlackBox}{\rule{1.5ex}{1.5ex}}  
\newcommand{\WhiteBox}{\rule{1.5ex}{1.5ex}}  

\newenvironment{proofblackdot}{\par\noindent{\it Proof.\
}}{\hfill\BlackBox\\[2mm]}

\newtheorem{theorem}{Theorem}
\newtheorem{lemma}{Lemma}

\newtheorem{corollary}{Corollary}
\newtheorem{definition}{Definition}

\title{Sufficient Conditions for Formation of a Network Topology by Self-interested Agents}
\begin{document}

\author{Swapnil Dhamal and Y. Narahari\\
\normalsize Indian Institute of Science, Bangalore, India}
\date{}

\maketitle
\begin{spacing}{1}

\begin{abstract} 
Networks such as organizational network of a global company play an important role in a variety of knowledge management and information diffusion tasks. The nodes in these networks correspond to individuals who are self-interested. The topology of these networks often plays a crucial role in deciding the ease and speed with which certain tasks can be accomplished using these networks. Consequently, growing a stable network having a certain topology is of interest. Motivated by this, we study the following important problem: given a certain desired network topology, under what conditions would  best response (link addition/deletion) strategies played by self-interested agents lead to formation of a pairwise stable network with only that topology. We study this interesting reverse engineering problem by proposing a natural model of recursive network formation. In this model, nodes enter the network sequentially and the utility of a node captures principal determinants of network formation, namely (1) benefits from immediate neighbors, (2) costs of maintaining links with immediate neighbors, (3) benefits from indirect neighbors, (4) bridging benefits, and (5) network entry fee. Based on this model, we analyze relevant network topologies such as star graph, complete graph, bipartite Tur\'an graph, and multiple stars with interconnected centers, and derive a set of sufficient conditions under which these topologies emerge as pairwise stable networks. We also study the social welfare properties of the above topologies.
\end{abstract}


\section{Introduction}
\label{sec:intro}

A primary reason for networks such as social networks to be formed is that every person or node gets certain benefits from the network and these benefits take different forms in different types of networks. However, these benefits do not come for free. Every node in the network has to pay 
a certain cost for maintaining links with its immediate neighbors or direct friends. 
This cost takes the form of time, money, or effort depending on the type of network. 
Owing to the tension between benefits and costs, self-interested or rational nodes 
think strategically while choosing their immediate neighbors. 
A stable network that forms out
of this process will have a topological structure as dictated by the individual
utilities and best response strategies of the nodes.

Often, stakeholders such as a social network owner or a social planner, who 
work with the networks so formed, would like the network to have a certain desirable topology
to facilitate efficient handling of knowledge management, information retrieval, and information
diffusion tasks using the network. Typical examples of these tasks
include enabling optimal communication among nodes for maximum
efficiency (knowledge management), extracting certain critical information from the nodes
(information retrieval), broadcasting some information to the nodes (information
diffusion), etc. If a particular topology is the most appropriate
for the set of tasks to be handled, it would be useful to orchestrate network formation in a way that
the required topology emerges as a stable network as a result of the network formation process.

A network in the current context can be naturally represented as a 
graph consisting of self-interested agents called {\em nodes} 
and connections or friendships called {\em links}. 
Our analysis in this paper is based on the equilibrium notion of {\em pairwise stability\/} which takes into account bilateral deviations arising from mutual agreement of link creation between two nodes, that Nash equilibrium fails to capture~\cite{jacksonbook}. Deletion is unilateral and a node can delete a link without consent from the other node. 
We recall the definition of pairwise stability
from the literature. 
Let $u_j(g)$ denote the utility that node $j$ gets when the network formed is $g$.

\begin{definition}\cite{jacksonbook}
A network is said to be pairwise stable if it is a best response for a node not to delete any of its links and there is no incentive for any two unconnected nodes to create a link between them. So $g$  is pairwise stable if \\(a) for each edge $e = (i, j) \in g$, $u_i(g \backslash \{e\}) \leq u_i(g)$ and  $u_j(g \backslash \{e\}) \leq u_j(g)$, and\\
(b) for each edge $e' = (i, j) \notin g$, if $ u_i(g \cup \{e'\})>u_i(g) $, then $u_j(g \cup {e'})<u_j(g)$.
\end{definition}

We also recall another important property, namely, efficiency.
\begin{definition}\cite{jacksonbook}
A network is said to be efficient if the sum of the utilities of the nodes in the network is maximal. So $g$ is efficient if it maximizes $\sum_{j\in N} u_j(g)$, that is, for all networks $g'$ on $N$, $\sum_{j\in N} u_j(g) \geq \sum_{j\in N} u_j(g') $.
\end{definition}

We consider that all nodes are homogeneous and they have global knowledge of the network.

\subsection{Motivation}
\label{sec:motiv}
One of the key problems addressed in the literature on social network formation is: 
given a set of self-interested nodes and a model of social network formation, 
which topologies would be  stable and which would be efficient.
The trade-off between stability and efficiency
is a key topic of  interest and concern in the literature on network formation.

\begin{figure}[!b]
\begin{tabular}{ccc}
\begin{minipage}{5cm}
\begin{center}
\includegraphics[scale=0.4]{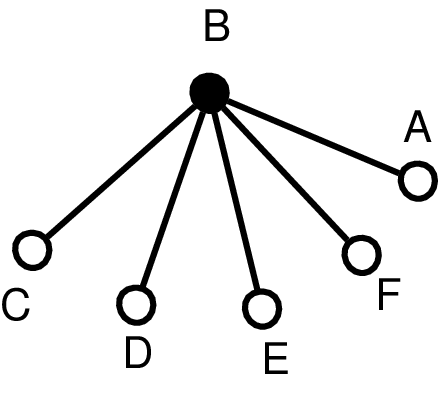}
\label{fig:motiv_star}
\end{center}
\end{minipage}
&
\begin{minipage}{5cm}
\begin{center}
\includegraphics[scale=0.4]{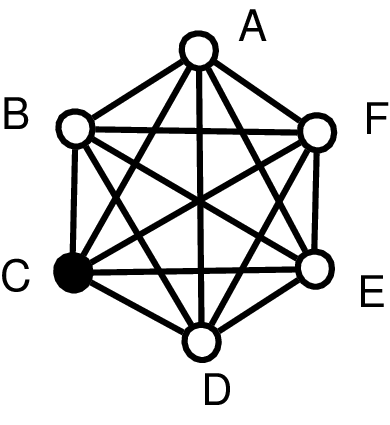}
\label{fig:motiv_complete}
\end{center}
\end{minipage}
&
\begin{minipage}{5cm}
\begin{center}
\includegraphics[scale=0.4]{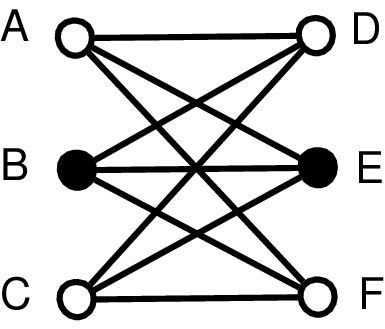}
\label{fig:motiv_bipartite}
\end{center}
\end{minipage}
\\
(a) Star & (b) Complete & (c) Bipartite Tur\'an
\end{tabular}

\begin{tabular}{cc}
\\
\begin{minipage}{7.8cm}
\begin{center}
\includegraphics[scale=0.4]{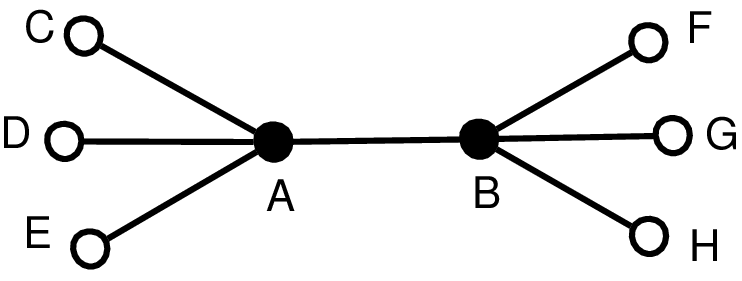}
\label{fig:motiv_2star}
\end{center}
\end{minipage}
&
\begin{minipage}{7.8cm}
\begin{center}
\includegraphics[scale=0.4]{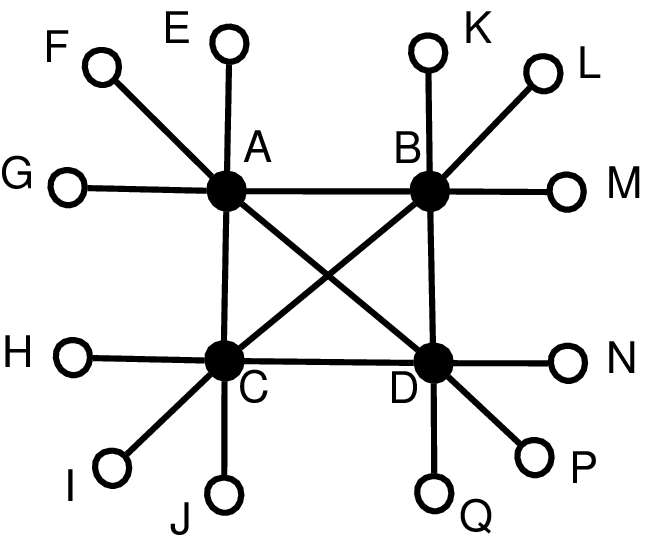}
\label{fig:motiv_kstar}
\end{center}
\end{minipage}
\\
(d) 2-star & (e) $k$-star ($k=4$)
\end{tabular}
\caption{Relevant topologies investigated in the paper}
\label{fig:motiv}
\end{figure}

In this paper, our focus is on the inverse problem, namely, 
given a certain desired network topology, under what conditions would  best response 
(link addition/deletion) strategies played 
by self-interested agents lead to formation of a stable (and perhaps efficient)
network with that topology. 
The problem becomes important because networks such as organizational network of a global company play an important role in a
variety of knowledge management, information retrieval, and information diffusion tasks. 
The topology
of these networks is one of the major factors that decides the ease and speed with which the 
above tasks can be accomplished. Often, a certain topology might serve the business interests
of the network owner better.
We explain this with some examples of relevant topologies shown
in Figure~\ref{fig:motiv}.

Consider a network where there is a need to rapidly spread some crucial information. 
The information may be received by any of the nodes and it is important that all other 
nodes also get the information at the earliest.
Also owing to the criticality of the information, it is desirable that there 
are redundant ways of communication, to take care of any link failures. 
In such cases, a complete network is ideal.

Consider a different scenario where the information is crucial, however 
there needs to be a moderator to verify the authenticity of the information before 
spreading it to the other nodes in the network (for example, it could be a rumor).
 Here a star network is desirable as the center could act as a moderator and any 
information that originates in any part of the network has to flow through the
moderator node before it can reach other nodes in the network. 
Virus inoculation is a related example where a star network is desirable 
since only the center needs to be inoculated in order to prevent spread of the virus
to other parts of the network, thus reducing the cost of inoculation.

Our next example concerns  two communities or clusters
 of a society where some or all members of a community
receive certain information simultaneously. The objective here is to forward
the information to the other community. Moreover, it is desirable to not have intra-section links to save on resources. In this case, it is desirable to have 
a bipartite network. Moreover, if the information is critical and urgent, 
requiring redundancy, a complete bipartite network is desirable. 
A bipartite Tur\'an network is a practical special case where both communities
are nearly of equal sizes.

Consider a generalization of the star network, where there are multiple centers and 
the leaf nodes are divided among the centers as evenly as possible. Such a network is desirable 
when the number of nodes is expected to be very large and there is a need for 
decentralization for efficiently controlling information in the network. 
We call such a network, $k$-star network (see Figure~\ref{fig:motiv}).

It is clear that depending on the tasks for which the network is used, a certain topology might
be better than the others. This provides the motivation for our work.

\subsection{Relevant Work}
\label{sec:relevant}

The modeling of strategic formation in a general network setting was first studied by Jackson and Wolinsky~\cite{jackson1996strategic}. This widely cited model, however, does not capture bridging benefits. Jackson~\cite{jackson2003stability} reviews several models of network formation in the literature and highlights that pairwise stable networks may not exist in some settings. 
Aumann and Myerson~\cite{myerson20} provide a sequential move game model where nodes are far-sighted, whereas Watts~\cite{watts618} considers a sequential move game model where nodes are myopic. In both of these approaches and in any sequential network formation model in general, the resulting network is based on the ordering in which links are altered and owing to random ordering, it is not clear which networks emerge.
Narayanam and Narahari~\cite{ramasuri1} investigate the topologies of networks formed with a generic model of network formation based on a value function, Myerson value.
Hummon~\cite{hummon2000utility} uses agent-based simulation approaches to explore the dynamics of network evolution based on the symmetric connections model~\cite{jackson1996strategic}. 
Goyal and Vega-Redondo~\cite{goyal2007structural} propose a non-cooperative game model capturing bridging benefits wherein they introduce the concept of {\em essential nodes}, which is a part of our utility model. 
Doreian~\cite{doreian2006actor}, given some conditions on a network, analytically arrives at specific networks that are pairwise stable. However, the complexity of analysis increases exponentially with the number of nodes and the analysis is limited to a network with only five nodes.
Some gaps in this analysis are addressed by Xie and Cui~\cite{xie2008cost},~\cite{xie2008note}.

The above models of social network formation assume that all nodes are present throughout the evolution of a network, which allows nodes to form links that may not be consistent with the desired network.
For instance, if our desired network is a star graph, with certain conditions on the network, a link between two nodes, of which one would play the role of the center, is desirable. However, with the same conditions, a link between other pairs is created with high probability, which is inconsistent with the desired star topology.
Furthermore, with all nodes present in an unorganized network, a random ordering over them in sequential network formation models adds to the complexity of analysis.
However, in most social networks, not all nodes are present from beginning itself. A network starts building up from a few nodes and gradually grows to its capacity. Our model captures such type of network formation.

There have been  a few approaches earlier to design incentives for nodes so that the resulting network is efficient.
Woodard and Parkes~\cite{woodard2003strategyproof} use mechanism design to design incentives so that the outcome is an efficient network. Mutuswami and Winter~\cite{mutuswami2002subscription} design a mechanism that ensures efficiency, budget balance, and equity. 
Though it is often assumed that the welfare of a network is based only on its efficiency, there are many
situations where this may not be true. A particular network
may not be efficient in itself, but it may be desirable for reasons
external to the network, as explained in Section~\ref{sec:motiv}.

\subsection{Contributions of the Paper}
\label{sec:gameinbrief}
In this paper, we study the inverse network formation problem, namely, 
under what conditions would a desired topology be obtained as a pairwise stable network
when self-interested agents form a network by playing best response strategies.  

\begin{itemize}
\item First we propose a recursive model of  network formation where nodes enter the
 network sequentially. The entry of a new node triggers an adjustment process in the network; this
adjustment process  continues until the network reaches a pairwise stable state. 
With this recursive model, we can guarantee that the network retains 
its topology in each of its stable states; also  the analysis 
can be carried out independent of the current number of nodes in the network.
The utility model we propose captures many key features: 
(a) benefits from immediate neighbors, (b) costs of maintaining links with immediate
neighbors, (c) benefits from indirect neighbors, (d) bridging benefits, 
and (e) an {\em entry fee\/}  for entering the network. 
\item With the above proposed model, we study common and important network topologies, 
namely, star graph, complete graph, bipartite Tur\'an graph, and  
$k$-star graph, 
and derive sufficient conditions under which a pairwise stable
network with the desired topology
will result under the proposed model of network formation. 
\item We also study the efficiency the above topologies.
\end{itemize}
To the best of our knowledge, this is the first detailed effort in investigating
the reverse engineering problem of obtaining a social network with desired topology.

\section{A Recursive Model of Network Formation}
\label{sec:model}

\begin{figure} [!t]
\begin{center}
\includegraphics[scale=0.5]{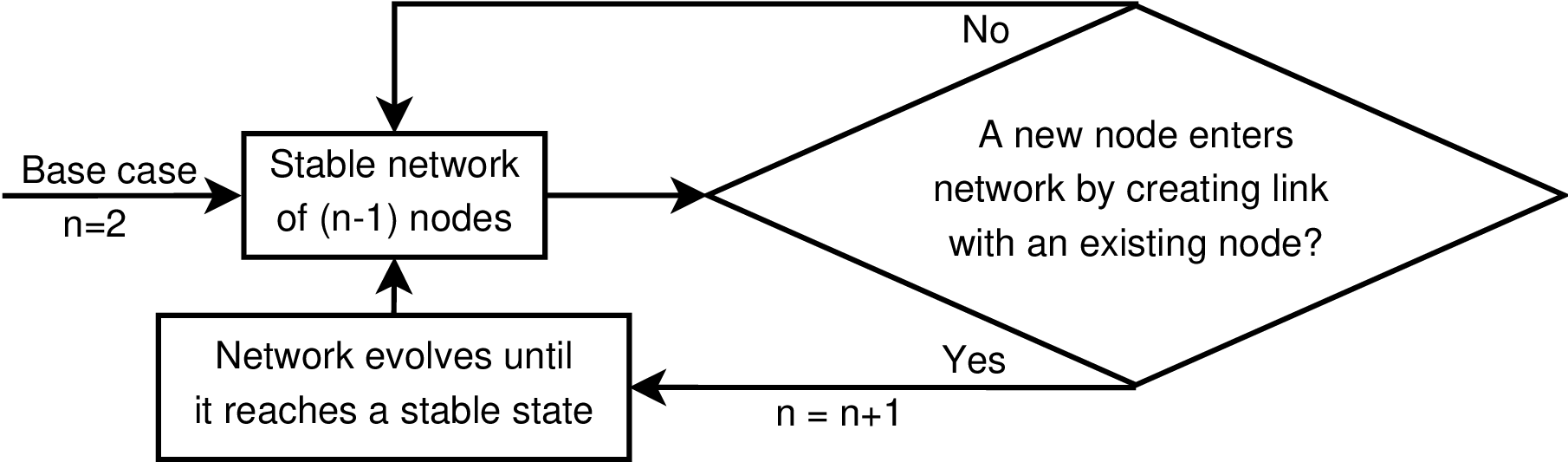}
\caption{Proposed model of network formation}
\label{fig:model}
\end{center}
\end{figure}

We consider that a game is played amongst a set of self-interested nodes, where the objective of each node is to maximize its profits or benefits that it gets from the network. The network consists of $n$ nodes at any given time, where $n$ could vary from one to a certain maximum number of nodes. 
The game starts with one node, whose only strategy is to remain in its current state. The strategy of the second node is to either (a) not enter the network or (b) form a link with the first node. We make an intuitive assumption that in order to be a part of the network, the second node has to propose a link with the first node and not vice versa. Also, for successful link creation, utility of the first node should not decrease. 
The network formed with these two nodes evolves to a pairwise stable network, which in the case of two nodes, does not result in any new network. After the network reaches a pairwise stable state, the third node considers entering the network. This process continues, which thus results in the formation of a pairwise stable network of $n$ nodes. We note that, in the process described above, 
no node in the network of $n-1$ nodes can create a link with the newly entering $n^{th}$ node until the latter successfully forms a link with one of the existing nodes in the network.
After the new node enters the network successfully, nodes who get to make their move are chosen at random at all time and the network evolves 
until it reaches a pairwise stable network consisting of $n$ nodes. Following this, a new ${(n+1)}^{th}$ node considers entering the network and the process goes on recursively. The assumption that a node considers entering the network only when it is pairwise stable might seem artificial in general social networks, but can be justified in organizational networks where entry of nodes can be controlled by a network administrator.
The model is depicted in Figure~\ref{fig:model}.

Each node has a set of strategies at any given time and it chooses its myopic best response strategy which maximizes its immediate utility. 
A strategy can be of one of the three types, namely (a) creating a link with a node that is not its immediate neighbor, (b) deleting a link with an immediate neighbor, or (c) maintaining status quo. 
Note that a node will compute whether a link it proposes decreases utility of the other node, because if it does, it is not its myopic best response as the link will not be accepted by the latter. Moreover, consistent with the notion of pairwise stability, if a node gets to make a move and altering a link does not strictly increase its utility, then it prefers not to alter it.

The proposed utility model, that we use for the purpose of our analysis, is described below.

\subsection{Utility Model}
\label{sec:utility}

As nodes have global knowledge of existing nodes in the network while making their decisions, for instance, creating a link with a faraway node, we propose a utility model that captures the global view of both indirect and bridging benefits. Our model takes the idea of {\em essential nodes} from the model proposed by Goyal and Vega-Redondo~\cite{goyal2007structural}.
A  node $j$ is said to be {\em essential} for $y$ and $z$ if $j$ lies on every path that joins $y$ and $z$ in the network. 
Whenever nodes $y$ and $z$ are directly connected, they get the entire benefits arising from the direct link. On the other hand, when they are indirectly connected with the help of other nodes, of which at least one is essential, $y$ and $z$ lose some fraction of the indirect benefits in the form of intermediation rents paid to the essential nodes without whom the communication is infeasible. 
Moreover, for simplicity of analysis, we assume that nodes that lie on path(s) connecting $y$ and $z$, but are not essential, do not get any share of the intermediation rents. So, when $y$ and $z$ are indirectly connected with the help of other nodes of which none is essential, they get the entire indirect benefits arising from their connection with each other.
In order to avoid discrete constraints on rents, such as summation of the fractions paid to be less than one, we assume that irrespective of the number of essential nodes connecting $y$ and $z$, they lose the same fraction.

We now describe the determinants of network formation that our model captures, and thus obtain expression for the utility function.
Table~\ref{tab:notation} enlists the notation we use in the rest of the paper.

\subsubsection{Network Entry Fee}
\label{sec:fee}
Since nodes enter a network one by one, we introduce a notion of {\em network entry fee}. 
This fee corresponds to some cost a node has to bear in order to be a part of the network. 
It is clear that, if a newly entering node wants its first connection to be with an existing node which is of high importance or degree, then it has to spend more time or effort. 
So the entry fee that the former pays is assumed to be an increasing function of the degree of the latter, say $d_{\text{T}}$. 
For simplicity of analysis, we assume the fee to be directly proportional to $d_{\text{T}}$ and call the proportionality constant, {\em network entry factor} $c_0$.

\begin{table}[t]
\begin{center}
  \begin{tabular}{| l | l |}

    \hline
\T \B $u_j$ & net utility that node $j$ gets from the network \\ \hline
\T \B $N$ & set of nodes present in the network \\ \hline
\T \B $d_j$ & degree of node $j$ \\ \hline
\T \B $b_i$ & benefits obtained from a node at distance $i$ in absence of rents \\ \hline
\T \B $c$ & costs incurred in maintaining link with an immediate neighbor \\ \hline
\T \B $l(j,w)$ & distance between nodes $j$ and $w$ \\ \hline
\T \B $\gamma$ & fraction of indirect benefits paid to the corresponding set of essential nodes \\ \hline
\T \B $E(j,w)$ & set of nodes essential to connect $j$ and $w$ \\ \hline
\T \B $e(j,w)$ & $|E(j,w)|$ \\ \hline
\T \B $c_0$ & network entry factor (see {\em Network Entry Fee}) \\ \hline
\T \B $\text{T}(j)$ & target node to which node $j$ connects to enter the network \\ \hline
\T \B $\textbf{I}_{\{j=\text{NE}\}}$ & 1 when $j$ is a newly entering node about to create its first link, else it is 0 \\ \hline

  \end{tabular}
\end{center}
\caption{Notation for the proposed utility model}
\label{tab:notation}
\end{table}

\subsubsection{Direct Benefits}
\label{sec:direct}
These benefits are obtained from immediate neighbors  in a network.
For a node $j$, these benefits equal $b_1$ times $d_j$.

\subsubsection{Link Costs}
\label{sec:cost}
These costs are the amount of resources like time, money, and effort a node has to spend in order to maintain links with its immediate neighbors.
For a node $j$, these costs equal $c$ times $d_j$.

\subsubsection{Indirect Benefits}
\label{sec:indirect}
These benefits are obtained from indirect neighbors or indirect friends and these decay with distance, that is $b_{i+1} < b_i$.
In the absence of rents, the total indirect benefits that a node $j$ gets is $\sum_{w \in N, \text{ } l(j,w)>1}{b_{l(j,w)}}$.

\subsubsection{Intermediation Rents}
\label{sec:rents}
Nodes pay a fraction $\gamma$ ($0 \leq \gamma < 1$) of the indirect benefits, in the form of additional favors or monetary transfers to the corresponding set of essential nodes, if any, and the loss incurred by a node $j$ due to these rents is $\sum_{w \in N , \text{ } E(j,w)\neq \phi}{\gamma b_{l(j,w)}}$.

\subsubsection{Bridging Benefits}
\label{sec:bridging}
In our model, a node gets bridging benefits for enabling communication between pairs of nodes which are otherwise disconnected.
Two nodes pay a fraction $\gamma$ of the indirect benefits to the set of essential nodes connecting them, which is assumed to be equally divided among the essential nodes connecting that pair. Let a node $j$ be one of the essential nodes connecting two nodes $y$ and $z$.
Both $y$ and $z$ benefit $b_{l(y,z)}$ each and so the connection produces a total benefit of $2 b_{l(y,z)}$. 
Each node from the set $E(y,z)$ gets a fraction $\frac{\gamma}{e(y,z)}$, the actual benefits being $\left( \frac{\gamma}{e(y,z)} \right) 2 b_{l(y,z)}$.
So the bridging benefits obtained by a node $j$ from the entire network is $\sum_{y,z \in N,\text{ }j \in E(y,z)}{ \left( \frac{\gamma}{e(y,z)} \right) 2 b_{l(y,z)} }$.

\subsubsection{Utility Function}
\label{sec:utilityfn}
For a node $j$, the utility function is a function of the network, that is $u_j:g \rightarrow \mathbb{R}$. We drop the notation $g$ from the following equation for readability.
Summing up all the determinants of network formation that our model captures, the utility function for node $j$ is given by

\begin{equation}
\label{eqn:utility}
u_j = -c_0d_{\text{T}(j)}\textbf{I}_{\{j=\text{NE}\}} + d_j(b_1-c) +\sum_{\substack{w \in N \\l(j,w)>1}}{b_{l(j,w)}} 
 - \sum_{\substack{w \in N \\E(j,w)\neq \phi}}{\gamma b_{l(j,w)}}  
    + \sum_{\substack{y,z \in N \\j \in E(y,z)}}{ \left( \frac{\gamma}{e(y,z)} \right) 2 b_{l(y,z)} } 
\end{equation}

\subsection{Dynamics of Network Formation}
\label{sec:directing}

The proposed model of network formation is based on a sequential move game and hence can be represented as an extensive form game tree. A snapshot of one such game tree is shown in Figure~\ref{fig:star}.

\subsubsection{Game Tree}
\label{sec:tree}

As the entry of each node in the network results in one game tree, the network formation process results in a series of game trees.
Each node of a game tree (not to be confused with a node of the network) represents a network state, while each branch represents a possible transition from a network state, owing to decision made by a node.
So, the root of a game tree represents the network state in which a new node is considering to enter the network.

A general way to find an equilibrium in an extensive form game is to use backward induction~\cite{osborne}. Our game is a special case of such a game where the players have bounded rationality, that is their best response strategies are myopic. So instead of the regular backward induction approach or the bottom-up approach, we take a top-down approach, which results in derivation of the same set of conditions under which a network topology gets formed. 
An {\em improving path} is a sequence of networks, where each transition is obtained by either two nodes choosing to add a link or one node choosing to delete a link. Thus, a pairwise stable network is one from which there is no improving path leaving it~\cite{jackson2002evolution}.
The notion of improving paths is a myopic one, and agents make their decisions of altering links without considering how their actions affect the decisions of other nodes and hence the evolution of network. Though this process of improving paths exhibits bounded rationality, it is a natural variation on best response dynamics and has some experimental justifications~\cite{pantz497}. 
So, deriving using the top-down approach leads to an intuitive understanding of the dynamics of network formation using the notion of improving paths.

\subsubsection{Notion of Types}
\label{sec:types}
As the order in which nodes take decision, is random, in a general game, the number of branches arising from each state in the game tree depends on the number of possible connections a node can be involved in (or number of possible connections with respect to a node). 
We say two nodes, say $A$ and $C$, of a graph $g$ are of the same type if there exists an automorphism
$f:V(g)\rightarrow V(g)$ such that $f(A)=C$, where $V(g)$ is the vertex set of $g$.
The implication of nodes being of the same type is that, for any automorphism $f$, if the best response strategy of node $A$ is to alter its link with node $D$, then the best response strategy of $f(A)$ is to alter its link with $f(D)$. So at any point in time, it is sufficient to consider the best response strategies of one node of each type.
We say two connections with respect to a node $B$, say with nodes $A$ and $C$, are of the same type if there exists an automorphism
$f$ such that $f(A)=C$ and $f(B)=B$.
The implication of connections being of the same type with respect to a node is that, the node is indifferent between the connections, irrespective of the utility model. Different types of connections with respect to a node form different branches in the game tree.

For example, in Figure~\ref{fig:motiv}(e), nodes $G$ and $H$ are of the same type. Also, the two possible connections $MG$ and $MH$ with respect to node $M$, are of the same type.
But the possible connections $EG$ and $EH$ with respect to node $E$, are not of the same type. So, these two strategies of node $E$, namely, connecting with nodes $G$ and $H$, form different branches in the game tree, implying that the utilities arising from these two types of connections are not necessarily equal.
In a network formation game with homogeneous nodes, the number of branches depends on the number of different types of possible connections with respect to a node, at that particular instant.
Furthermore, as social networks inherently have low diameter~\cite{milgram1967small} and we are primarily interested in the formation of special topologies in a recursive manner (nodes are already organized according to the topology and the objective is to extend the topology to one more node, so the existing nodes play the same role as before, and most or all of the existing links do not change), the number of different types of nodes, as well as the number of different types of possible connections with respect to a node, at any instant, is a small constant, thus simplifying the analysis.

\begin{figure}[!b]

\begin{tabular}{cc}
\begin{minipage}{4.5cm}
\begin{center}
\vspace{-2.5in}
\includegraphics[scale=0.42]{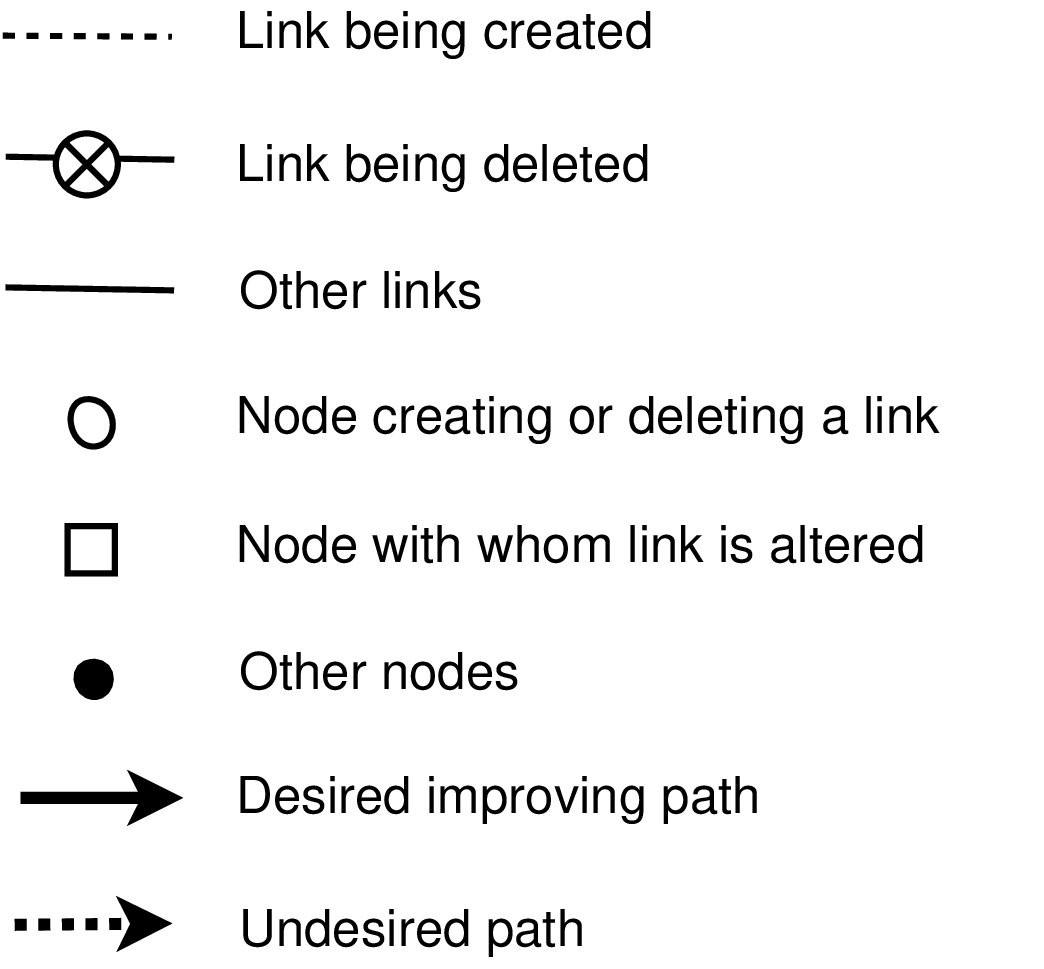}
\end{center}
\end{minipage}
&
\begin{minipage}{5.5cm}
\begin{center}
\includegraphics[scale=0.42]{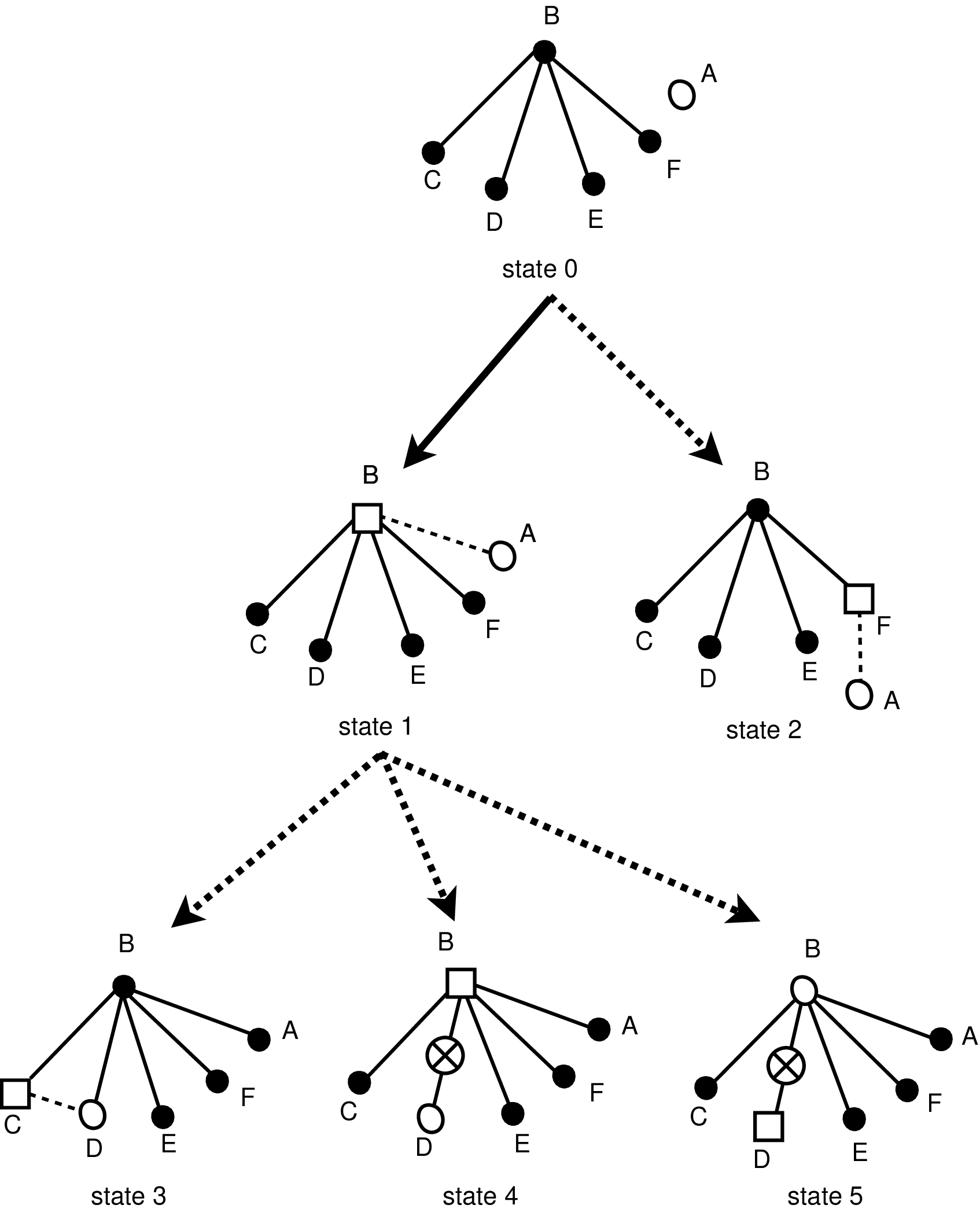}
\end{center}
\end{minipage}
\end{tabular}
\caption{Directing  the Dynamics of Star Formation}
\label{fig:star}

\end{figure}

\subsubsection{Directing the Dynamics}
\label{directing_dymanics}

The procedure for deriving sufficient conditions for the formation of a given topology is similar to {\em mathematical induction}. Consider a base case network with very few nodes (two in our analysis).
We derive conditions so that the network formed with these few nodes has the desired topology. Then using induction, we assume that a network with $n-1$ nodes has the desired topology, and derive conditions so that, the network with $n$ nodes, also has that topology. 
Without loss of generality, we explain this procedure with the example of star topology, referring to Figure~\ref{fig:star}. We derive conditions for base case in the proof of Theorem~\ref{thm:star}. Now assuming that the network formed with $n-1$ nodes is a star, we derive conditions so that the network of $n$ nodes is also a star.

In Figure~\ref{fig:star}, node $A$ is the newly entering node and the network is currently in state 0 where a star with $n-1$ nodes is already formed. 
Note that in state 0, with respect to node $A$, there are two types of possible connections: (a) with the center and (b) with a leaf node. In states 1, 3, 4 and 5, there are two types of nodes, and two types of possible connections with respect to a leaf node and one with respect to the center. It will be seen that, the network is directed to not enter state 2, so even though there are four different types of nodes, it is not a matter of concern. 
Firstly, we want node $A$ to connect to the center and enter state 1 by choosing the improving path that transits the network from state 0 to state 1 in Figure~\ref{fig:star}. So utility of node $A$ in state 1 should be greater than that in state 0, that is $u_A(1) > u_A(0)$, where $u$ is any utility function, Equation~(\ref{eqn:utility}) being one such function. Similarly, for node $B$ to accept the link from node $A$, $B$'s utility should not decrease, that is $u_B(1) \geq u_B(0)$. 
Also we do not want node $A$ to connect to any of the leaf nodes, that is, we do not want the network to enter state 2. Note that there might be an improving path from state 2 that eventually results in a star, but as we are concerned with sufficient conditions, we discard state 2 in order to shorten the analysis. One way to ensure that the network does not enter state 2, irrespective of whether it lies on an improving path, is by making it less favorable for node $A$ than the desired state 1. That is, starting from state 0, there exists an alternative improving path which gives node $A$ better utility than entering state 2. We say that, for node $A$, the strategy of creating a link with any of the leaf nodes is dominated by the strategy of creating link with the center. For state 2 to be dominated by state 1, $u_A(2) < u_A(1)$. Another way to ensure the same is by having a condition for a leaf node such that, accepting a link from node $A$ decreases its utility and so a leaf node does not accept the link, thus forcing node $A$ to connect to the center, that is $u_j(2) < u_j(0)$ for $j=C,D,E,F$. So the network enters state 1, which is our desired state.

In order to ensure pairwise stability of our desired state, there should be no improving paths leading out of it, for which we need to take care of two cases. First, when node $B$ gets to make its move, it can either (a) break any of its links and enter state 5 or (b) remain in the same state 1. 
As we do not want the network to enter state 5, node $B$'s utility should be at least as good remaining in state 1 as entering 5, that is $u_B(1) \geq u_B(5)$.
Second, when any of the leaf nodes is chosen at random, it can either (a) create a link with some other leaf node and enter state 3 or (b) delete its link with the center and enter state 4, or (c) remain in the same state 1. As we want the network to stay in state 1, the additional conditions respectively are $u_j(1) \geq u_j(3)$ and $u_j(1) \geq u_j(4)$ for $j=A,C,D,E,F$.
Thus we direct the dynamics of network formation along a desired improving path by imposing a set of conditions ensuring that the resulting network is in the desired state or has the desired topology. In the following section, we analyze some common network topologies using our model of recursive and sequential network formation.

\section{Sufficient Conditions for Formation of Relevant Topologies}
\label{sec:analysis}

In this section, we analyze the dynamics of formation of several relevant network topologies, namely star, complete graph, bipartite Tur\'an graph, 2-star, and $k$-star, and derive sufficient conditions for their formation.
Note that the conditions derived for any particular network topology are sufficient under the given setting, and there may exist alternative conditions that result in the same topology. 
We use Equation~(\ref{eqn:utility}) for mathematically deriving the conditions.

\begin{theorem}
\label{thm:star}
For a given network, if $b_1-b_2 + \gamma b_2 \leq c < b_1$ and $c_0 < \left( 1-\gamma \right) \left( b_2-b_3 \right)$, 
the resulting topology is a star.
\end{theorem}

\begin{proofblackdot}
Refer to Figure~\ref{fig:star} throughout the proof. For the base case of $n=2$, the requirement for the second node to connect 
to the first is that its utility should become strictly positive. Also as the first node has degree $0$, there is no entry fee.
\begin{equation}
\label{B1for2} 
0 < b_1-c  \Leftrightarrow c < b_1
\end{equation}
Now, consider a star consisting of $n-1$ nodes. Let the newly entering $n^{th}$ node get to make a decision of whether to enter the network. For $n\geq3$, if the entering node connects to the center, it gets indirect benefits of $b_2$ each from $n-2$ nodes. But as the center is essential in order to connect the newly entering node with the other leaf nodes, the new node has to pay $\gamma$ fraction of these benefits to the center. Also, it has to pay an entry fee of $(n-2)c_0$ as the degree of center is $n-2$. So in Figure~\ref{fig:star}, 
$u_A(0) < u_A(1)$ gives
\begin{equation}
\nonumber
0 < b_1-c+(n-2) \left( 1-\gamma \right) b_2-(n-2)c_0
\end{equation}
\begin{equation}
\nonumber
\Leftrightarrow  c < b_1+(n-2) \left( \left( 1-\gamma \right) b_2-c_0 \right)
\end{equation}
As it needs to be true for all $n \geq 3$, we set the condition to
\begin{equation}
\nonumber
 c < \min_{n \geq 3} \Big\{ b_1+(n-2) \left( \left( 1-\gamma \right) b_2-c_0 \right) \Big\}
\end{equation}
\begin{equation}
\label{B1}
\Leftarrow  c <b_1+  \left( 1-\gamma \right)  b_2-c_0
\end{equation}
The last step is obtained so that the condition for link cost is independent of the upper limit on the number of nodes, by enforcing
\begin{equation}
\label{B1forc0}
c_0 \leq  \left( 1-\gamma \right) b_2
\end{equation}
which enables us to substitute $n=3$ and the condition holds for all $n\geq 3$.\\
For the center to accept a link from the newly entering node, we need to have $u_B(0) \leq u_B(1)$.
For $n=2$, the requirement for the first node to accept link from the second node is $0 \leq b_1-c$ which is satisfied by Inequality~(\ref{B1for2}).
For $n=3$, as the center is essential for connecting the other two nodes separated by distance two, it gets $\gamma$ fraction of $b_2$ from both the nodes. So it gets bridging benefits of $2 \gamma b_2$.
\begin{equation}
\nonumber
 b_1-c \leq 2(b_1-c)+2\gamma b_2
\end{equation}
\begin{equation}
\nonumber
\Leftrightarrow c \leq b_1+ 2\gamma b_2
\end{equation}
This condition is satisfied by Inequality~(\ref{B1for2}).
For $n\geq 4$, prior to entry of the new node, the center alone connected \tiny{$\dbinom{n-2}{2}$} 
\normalsize pairs of nodes at distance two from each other, while after connecting with the new node, the center is the sole connection for \tiny{$\dbinom{n-1}{2}$} 
\normalsize such pairs.
So the required condition:
\begin{equation}
\nonumber
 (n-2)(b_1-c)+\gamma \dbinom{n-2}{2}2 b_2 \leq (n-1)(b_1-c)+\gamma \dbinom{n-1}{2}2 b_2
\end{equation}
This condition is satisfied by Inequality~(\ref{B1for2}) for all $n \geq 4$.\\
For the newly entering node to prefer the center over a leaf node as its first connection (not applicable for $n=2$ and $3$), we need $u_A(1) > u_A(2)$.
\begin{equation}
\nonumber
 b_1-c +(n-2) \left( 1-\gamma \right) b_2 -(n-2)c_0  > b_1-c+ \left( 1-\gamma \right) b_2  +(n-3) \left( 1-\gamma \right) b_3-c_0
\end{equation}
\begin{equation}
\label{B3a}
\Leftrightarrow c_0 <  \left( 1-\gamma \right) \left( b_2-b_3 \right)
\end{equation}
Alternatively, the newly entering node may want to connect to the leaf node, but the leaf node's utility decreases. In that case, the alternative condition can be $u_j(2)<u_j(0)$ for $j=C,D,E,F$.
Note that this leaf node gets bridging benefits of $2\gamma b_2$ for being essential in connecting the new node with the center. Also, as it is one of the two essential nodes in connecting the new node with the other $n-3$ leaf nodes (the other being the center), it gets bridging benefits of $(n-3) (\frac{\gamma}{2})2 b_3 = (n-3) \gamma b_3$.
\begin{equation}
\nonumber
b_1-c +(n-3) \left( 1-\gamma \right) b_2 > 2(b_1-c)+(n-3) \left( 1-\gamma \right) b_2 + 2\gamma b_2 + (n-3) \gamma b_3
\end{equation}
which gives $c>b_1+ 2\gamma b_2 + (n-3) \gamma b_3$. But this is inconsistent with the condition in Inequality~(\ref{B1for2}). So in order to ensure that the newly entering node connects to the center and not to any of the leaf nodes, we use Inequality~(\ref{B3a}) only.

Now that a star of $n$ nodes is formed, we ensure its pairwise stability by deriving conditions for the same. 
Firstly, we ensure that the center does not delete any of its links. So we need $u_B(1) \geq u_B(5)$. Note that from the center's point of view, state $5$ is same as state $0$ and as we have seen earlier that $u_B(0) \leq u_B(1)$, the required condition $u_B(4) \leq u_B(1)$ is already ensured.\\
Next, no two leaf nodes should form a link between them. So we should ensure that, not creating a link between them is at least as good for them as creating, that is $u_j(1) \geq u_j(3)$ for any leaf node $j$. This condition is applicable for $n\geq 3$. 
\begin{equation}
\nonumber
b_1-c+(n-2) \left( 1-\gamma \right) b_2 \geq 2(b_1-c)+(n-3) \left( 1-\gamma \right) b_2
\end{equation}
\begin{equation}
\label{B4}
\Leftrightarrow c \geq b_1-b_2+\gamma b_2
\end{equation}
For a leaf node to not delete its link with the center, we need $u_j(1) \geq u_j(4)$ for any leaf node $j$. For $n \geq 2$, we have
\begin{equation}
\nonumber
b_1-c+(n-2)  \left( 1-\gamma \right) b_2  \geq 0
\end{equation}
\begin{equation}
\nonumber
\Leftrightarrow c \leq b_1+(n-2)  \left( 1-\gamma \right) b_2
\end{equation}
which is a weaker condition than Inequality~(\ref{B1for2}) for $n\geq 2$.

Note that Inequalities~(\ref{B1for2}) and (\ref{B3a}) put together are stronger than Inequalities~(\ref{B1}) and (\ref{B1forc0}) combined. 
We get the required result using Inequalities~(\ref{B1for2}), (\ref{B3a}) and (\ref{B4}).
\end{proofblackdot}

We provide proof of Theorem~\ref{thm:bipartite} in the Appendix. Proofs of Theorems~\ref{thm:complete} and \ref{thm:2star} use similar technique.

\begin{theorem}
\label{thm:complete}
For a network, if $c < b_1-b_2$ and $c_0 \leq \left( 1-\gamma \right) b_2$, the resulting topology is a complete graph.
\end{theorem}

\begin{theorem}
\label{thm:bipartite}
For a network with $\gamma <   \frac{b_2 - b_3}{3b_2 - b_3} $, if $b_1-b_2+ \gamma \left( 3b_2 - b_3 \right) <  c < b_1 - b_3$ 
and $\left( 1-\gamma \right) \left( b_2-b_3 \right) < c_0 \leq \left( 1-\gamma \right) b_2$, the resulting topology is a 
bipartite Tur\'an graph.
\end{theorem}

\begin{theorem}
\label{thm:2star}
Let $\sigma$ be the upper bound on the number of nodes that can enter the network and $\lambda = \lceil \frac{\sigma}{2} -1 \rceil \left( 2b_2-b_3 \right)$.
Then, if $\left( 1-\gamma \right) \left( b_2-b_3 \right) < c_0 < \left( 1-\gamma \right) \left( b_2-b_4 \right) $ and either \\
(i) $ \frac{b_2-b_3}{\lambda -b_3} \leq \gamma < \frac{b_2}{\lambda +1}$ and $b_1-b_2+ \gamma b_2 + \gamma \lambda <  c < b_1$, or\\
 (ii) $\gamma <   \min \Big\{ \frac{b_2}{\lambda +1} , \frac{b_2-b_3}{\lambda -b_3} \Big\}$ and $b_1-b_3+ \gamma \left( b_2 + b_3 \right) \leq  c < b_1$,\\
the resulting topology is a 
2-star.
\end{theorem}

For arbitrarily large upper bound on the number of nodes that can enter the network, the following corollary is immediate from the above theorem.

\begin{corollary}
\label{cor:2star}
For a network with $\gamma=0$, if $b_1-b_3 \leq  c < b_1$ and $b_2-b_3< c_0 < b_2-b_4$,
the resulting topology is a 
2-star.
\end{corollary}

\subsection{Base Graph}
\label{sec:base}

The sufficient conditions for the formation of topologies analyzed so far, are obtained, starting from the base graph or the starting graph consisting of a single node (corresponding to the base case of formation of a network with $n=2$).
However, in case of some topologies, under a given utility model, the conditions required for its formation on discretely small number of nodes, are inconsistent with that required on arbitrarily large number of nodes. Lemma~\ref{lem:kstar} shows that $k$-star is one such topology.
We provide its proof in the Appendix.

\begin{lemma}
\label{lem:kstar0}
Under the proposed utility model, for the entire family of $k$-star networks ($k\geq 3$) to be pairwise stable, it is necessary that 
$\gamma=0$ and $c=b_1-b_3$.
\end{lemma}

\begin{proofblackdot}
We consider two scenarios sufficient to prove this.

\noindent
I) No center should delete its link with any other center: Here, only one case is enough to be considered, that is, when each center has just one leaf node 
since in all other cases, the benefits obtained by each center from the connection with other centers is at least as much. For $k=3$, \vspace{-.2cm} 
\begin{equation}
\nonumber
\begin{split}
3(b_1-c) + 2(1-\gamma)b_2 + \gamma(1)(2)2b_2 + \frac{\gamma}{2}(1)(2)b_3
 \geq 2(b_1-c) + (1-\gamma)b_2 + (1-\gamma)b_2 \\+ (1-\gamma)b_3 + \gamma(1)(1)2b_2 + \frac{\gamma}{2}(1)(1)2b_3 \\+ \frac{\gamma}{2}(1)(1)2b_3 + \frac{\gamma}{3}(1)(1)2b_4
\end{split}
\end{equation}
which gives
\begin{equation}
\label{eq:kstarineq1}
c \leq b_1-b_3+\gamma(2b_2+b_3)-\frac{2\gamma}{3}b_4
\end{equation}
For $k\geq 4$,
\begin{equation}
\nonumber
\begin{split}
(k-1+1)(b_1-c) + (k-1)(1-\gamma)b_2 + \gamma(1)(k-1)2b_2 + \frac{\gamma}{2}(1)(k-1)2b_3\\
 \geq (k-2+1)(b_1-c) + (k-2)(1-\gamma)b_2 + b_2 + (1-\gamma)b_3 + \gamma(1)(k-2)2b_2\\ + \frac{\gamma}{2}(1)(k-2)2b_3 + \gamma(1)(1)2b_3 + \frac{\gamma}{2}(1)(1)2b_4
\end{split}
\end{equation}
which gives
\begin{equation}
\label{eq:kstarineq2}
c \leq b_1-b_3+\gamma(b_2-b_4)
\end{equation}

\noindent
II) Leaf nodes of different centers should not form a link with each other: Consider a leaf node. Let $m_i$ be the number of leaf nodes connected to the center to which the leaf node under consideration is connected. For $k\geq 3$,
\begin{equation}
\nonumber
\begin{split}
2(b_1-c) + (m_i-1)(1-\gamma)b_2 + (1-\gamma)b_3 (\sum_{j \neq i}m_i - 1) + (k-1)b_2 \\
\leq b_1-c + (m_i-1)(1-\gamma)b_2 + (1-\gamma)b_3 \sum_{j \neq i}m_i + (k-1)(1-\gamma)b_2
\end{split}
\end{equation}
which gives
\begin{equation}
\label{eq:kstarineq3}
c \geq b_1-b_3+\gamma((k-1)b_2+b_3)
\end{equation}
The only way to satisfy Inequalities~(\ref{eq:kstarineq1}), (\ref{eq:kstarineq2}) and (\ref{eq:kstarineq3}) simultaneously is by setting 
\begin{equation}
\label{eq:kstargamma}
\gamma=0
\end{equation}
and
\begin{equation}
\label{eq:kstarcost}
c=b_1-b_3
\end{equation}
\end{proofblackdot}

\begin{lemma}
\label{lem:kstar}
Under the proposed network formation and utility models, starting with a network consisting of a single node, $k$-star network ($k\geq3$) cannot be formed.
\end{lemma}

\begin{proofblackdot}
From Lemma~\ref{lem:kstar0}, the conditions necessary for the family of $k$-star networks to be pairwise stable are $\gamma=0$ and $c=b_1-b_3$, which are sufficient conditions for the formation of 2-star network according to Corollary~\ref{cor:2star}.
\end{proofblackdot}

The following lemma shows a result for a more general class of network formation models with some restriction on the desired improving path.

\begin{lemma}
\label{lem:kstar2}
Under the proposed utility model and the class of evolution models that allow 
altering at most one link at a time based on myopic best response strategies, starting with a network consisting of a single node,
one cannot find
sufficient conditions for the formation of a $k$-star network ($k\geq3$) by directing the evolution of a network along a shortest improving path (where each network is obtained from the preceding network by adding a link).
\end{lemma}
\begin{proofblackdot}
It is clear from Lemma~\ref{lem:kstar0} that $\gamma=0$ and $c=b_1-b_3$ are necessary for the family of $k$-star networks ($k\geq 3$) to be pairwise stable. We consider three cases how a $k$-star network builds.

Case (i) The first $k$ nodes form a complete network amongst themselves first, following which, other nodes enter, who then play the role of leaf nodes: Lemma~\ref{lem:kstar} shows that a $k$-star network cannot be formed if the topology is to be maintained in each pairwise stable state. So there exist no conditions under which a $k$-star network will be formed.
In the rest of the cases, the centers are connected to some leaf nodes and there will eventually come a time when two centers at distance two from each other, have to form a mutual link.

Case (ii) At least one of these two centers have no leaf node: It can be seen that, with the necessary condition $\gamma=0$, if $c \geq b_1-b_2$, no node would want to create a link with another node which is at distance two or more from it, if all of latter's direct neighbors, if any, are direct neighbors of the former. So creating a link with the center (at distance two from it) having no leaf is not beneficial for the other center as $c=b_1-b_3$.

Case (iii) Both centers have at least one leaf node each: It can be seen that if $c< b_1-b_4$, nodes who are at distance four or more benefit by creating a link between them. As the leaf nodes of these two centers are at distance four from each other, it is desirable for them to create link between them as $c=b_1-b_3$.
\end{proofblackdot}

A possible and reasonable solution to overcome this problem is to analyze the network formation process, starting from some other base graph. The base graph can be obtained by some other method, one of which could be providing additional incentives to the nodes of the base graph.
For instance, for analyzing the formation of $k$-star, the base graph is taken to be the complete network on the $k$ centers, with the centers connecting to one leaf node each. As the base graph consists of $2k$ nodes, the induction starts with the base case for formation of $k$-star network with $n=2k+1$.
Theorem~\ref{thm:kstar} gives the sufficient conditions for the formation of $k$-star network provided the network starts building itself from this base graph. Its proof uses technique similar to that of Theorem~\ref{thm:bipartite}.

\begin{theorem}
\label{thm:kstar}
For a network starting with the base graph for $k$-star ($k \geq 3$), and $\gamma =0 $, if $c =b_1-b_3 $ 
and $ b_2-b_3  < c_0 < b_2-b_4$, the resulting topology is a 
$k$-star.
\end{theorem}

\begin{proofblackdot}
It is clear from Lemma~\ref{lem:kstar0} that under the proposed utility model, for the family of $k$-star networks ($k\geq 3$) to be pairwise stable, it is necessary that $\gamma=0$, that is, we need $\gamma=0$ in order to arrive at all possible $k$-star networks for a given $k$, and hence forms the necessary part of sufficient conditions for the formation of a $k$-star network. Hence, for the rest of this proof, we will assume $\gamma=0$.

 Without loss of generality, assume some indexing over the $k$ centers from $1$ to $k$ such that any center connected to more leaf nodes has a higher index than any center connected to less leaf nodes. 
Let $C_i$ be the center with index $i$ and $m_i$ be the number of leaf nodes it is connected to. 
Also we start with a base graph in which every center is connected to one leaf node and the number of leaf nodes connected to each center increases as the process goes on. 
So we have, $1 \leq m_1 \leq m_2 \leq \dots \leq m_k$.\\

\noindent
\textbf{For the newly entering node to propose entering the network:}
Our objective is to ensure that the newly entering node connects to a center with the least number of leaf nodes in order to maintain balance over the number of leaf nodes connected to the centers. Without loss of generality, assume that we want the newly entering node to connect to $C_1$. The utility of the newly entering node should be positive after doing so.
\begin{equation}
\nonumber
b_1 - c + (m_1+k-1)\left(b_2-c_0\right) +b_3\sum_{i=2}^{k} m_i > 0
\end{equation}
Since the minimum value of $m_i$ is $1$ for any $i$, the above condition is true if
\begin{equation}
\nonumber
c < b_1 + k \left(b_2-c_0\right) + (k-1)b_3
\end{equation}
This is true if 
\begin{equation}
\nonumber
c<b_1 \text{ (satisfied by Equation~(\ref{eq:kstarcost}))}
\end{equation}
and
\begin{equation}
\label{G1forc0}
c_0 \leq b_2
\end{equation}

\noindent
\textbf{The newly entering node should connect to a center with the least number of leaf nodes, whenever applicable:}
This case does not arise when all centers have the same number of leaf nodes. Moreover, the way we direct the evolution of the network, the number of leaf nodes connected to any two centers differs by at most one. Without loss of generality, assume that we want the newly entering node to connect to $C_1$. Consider a center $C_p$ such that $m_p = m_1+1$. So the newly entering node should prefer connecting to $C_1$ over connecting to $C_p$.
\begin{equation}
\nonumber
b_1-c + (m_1+k-1) \left(b_2 - c_0 \right) +b_3\sum_{i=2}^{k} m_i 
> b_1-c + (m_p+k-1) \left( b_2 - c_0 \right) + b_3\sum_{\substack{1 \leq i \leq k \\ i \neq p}} m_i
\end{equation}
As $m_p = m_1 + 1$, we have
\begin{equation}
\label{G3}
c_0 > b_2-b_3
\end{equation}

\noindent
\textbf{For a center with the least number of leaf nodes to accept the link from the newly entering node:}
It can be easily seen that this is ensured by Equation~(\ref{eq:kstarcost}).\\

\noindent
\textbf{The newly entering node should not connect to any leaf node:}
It can be easily seen that owing to benefits degrading with distance, for the newly entering node, connecting to any leaf node which is connected to a center with the most number of leaf nodes strictly dominates connecting to any other leaf node, whenever applicable. So it is sufficient to ensure that the newly entering node does not connect to any leaf node which is connected to a center with the most number of leaf nodes.
This can be done by ensuring that for the newly entering node, connecting to a center with the least number of leaf nodes strictly dominates connecting to any leaf node which is connected to a center with the most number of leaf nodes.
\begin{equation}
\nonumber
b_1-c + (m_1+k-1)(b_2-c_0) +  b_3\sum_{i=2}^k m_i 
> b_1-c +b_2 - c_0 + (m_k + k-2)b_3 +  b_4\sum_{i=1}^{k-1} m_i 
\end{equation}
We need to consider two cases (i) $m_k = m_1+1$ and (ii) $m_k=m_1$\\
Case (i) $m_k = m_1+1$: Substituting  this value of $m_k$ gives
\begin{equation}
\nonumber
(m_1+k-1)(b_2-c_0) +  (b_3-b_4)\sum_{i=2}^{k-1} m_i + m_1 (b_3-b_4)+ b_3
> b_2 - c_0 + (m_1 + k-1)b_3 
\end{equation}
As the minimum value of $\sum_{i=2}^{k-1}m_i$ is $k-2$, the above remains true if we replace $\sum_{i=2}^{k-1}m_i$ by $k-2$. Further simplification gives
\begin{equation}
\nonumber
(m_1+k-2)(b_2-b_4-c_0)>0
\end{equation}
Since $m_1+k-2>0$ is positive, we must have
\begin{equation}
\label{G4forc0}
c_0 < b_2-b_4
\end{equation}
Case (ii) $m_k=m_1$: It can be similarly shown that Equation~(\ref{G4forc0}) is the sufficient condition.\\

\noindent
Now that the newly entering node enters in a way such that $k$-star network is formed, we have to ensure that no further modifications of links occur so that the network formed thus is pairwise stable.\\

\noindent
\textbf{For centers and the corresponding leaf nodes to not delete the link between them:} It can be easily seen that $c<b_1$, a weaker condition than Equation~(\ref{eq:kstarcost}), is a sufficient condition to ensure this.\\

\noindent
\textbf{No center should delete its link with any other center:} This is ensured by Inequalities~(\ref{eq:kstarineq1}) and (\ref{eq:kstarineq2}) as explained in Lemma~\ref{lem:kstar0}. \\

\noindent
\textbf{Leaf nodes of a center should not form a link with each other:} The net benefit that a leaf node gets by forming such a link should be non-positive.
\begin{equation}
\nonumber
b_1-c - b_2 \leq 0
\end{equation}
\begin{equation}
\nonumber
\Leftrightarrow c \geq b_1-b_2
\end{equation}
which is satisfied by Equation~(\ref{eq:kstarcost}).\\

\noindent
\textbf{Leaf nodes of different centers should not form a link with each other:} This is ensured by Inequality~(\ref{eq:kstarineq3}) as explained in Lemma~\ref{lem:kstar0}. \\

\noindent
\textbf{Link between a center and leaf node of other centers should not be created:} Let $C_i$ be the center under consideration and the leaf node under consideration be connected to $C_j$ ($j\neq i$). There are two ways to ensure this. First is to ensure that a center neither proposes nor accepts a link with a leaf node of other centers. This mathematically is
\begin{equation}
\nonumber
(k-1+m_i)(b_1-c) + b_2 \sum_{k \neq i}m_k > (k-1+m_i+1)(b_1-c)+b_2(\sum_{k \neq i}m_k-1)
\end{equation}
\begin{equation}
\nonumber
\Leftrightarrow c > b_1-b_2
\end{equation}

An alternative to this condition is to ensure that a leaf node neither proposes nor accepts a link with a center to which it is not connected, but since this condition is already satisfied by Equation~(\ref{eq:kstarcost}), this alternative need not be considered.

Equations~(\ref{eq:kstargamma}), (\ref{eq:kstarcost}), (\ref{G1forc0}), (\ref{G3}) and (\ref{G4forc0}) give the required sufficient conditions for the formation of a $k$-star network.
\end{proofblackdot}

\subsection{Intuition Behind the Results}
\label{sec:explain}

The network entry fee has an impact on the resulting topology as seen from the above theorems. For instance, in Theorems~\ref{thm:star} and \ref{thm:bipartite}, the intervals spanned by the values of $c$ and $\gamma$ may intersect, but the values of network entry factor $c_0$ span mutually exclusive intervals separated at $(1-\gamma)(b_2-b_3)$. In case of star, $c_0$ is low and a newly entering node can afford to connect to the center, which in general, has very high degree. In case of bipartite Tur\'an graph, it is important to ensure that the sizes of the two partitions are as equal as possible. As $c_0$ is high, a newly entering node connects to a node with a lower degree, whenever applicable, that is, to a node that belongs to the partition with more number of nodes, and hence the newly entering node potentially becomes a part of the partition with fewer number of nodes, thus maintaining a balance between sizes of the two partitions. 
In case of $k$-star, the objective is to ensure that a newly entering node connects to a node with moderate degree, that is, the network entry factor is not so high that a newly entering node prefers connecting to a leaf node and also not so low that it prefers connecting to a center with the highest degree. This intuition is clearly reflected in Theorems~\ref{thm:2star} and \ref{thm:kstar} where $c_0$ takes intermediate values.
In general, a high value of {\em network entry factor} $c_0$ lays the foundation for formation of a regular graph. 

It is clear that a complete network is formed when the costs of maintaining links is extremely low, as reflected in Theorem~\ref{thm:complete}. The remaining topologies are formed in the intermediate ranges of $c$. From Theorems~\ref{thm:bipartite}, \ref{thm:2star} and \ref{thm:kstar}, it can be seen that the feasibility of a network being formed depends on the values of $\gamma$ also, which arises owing to contrasting densities of connections in a network. 
For instance, in a bipartite Tur\'an network, nodes from different partitions are densely connected with each other, while that from the same partition are not connected at all. Similarly, in a $k$-star network, there is an extreme contrast in the densities of connections (dense amongst centers and sparse for leaf nodes).

\subsection{Connection to Efficiency}
\label{sec:efficiency}

In this section, we analyze the efficiency of the concerned networks. 
The conditions are sufficient and not necessary, and so there may exist other sets of conditions that result in a given topology. We analyze the efficiency based on the assumption that the networks are formed using the derived conditions.

From Equation~(\ref{eqn:utility}), the intermediation rents are transferable among the nodes, and so do not affect the efficiency of a network. Furthermore, the network entry fee is paid by any node at most once, and so does not account for efficiency in the long run. So the expression for efficiency of a network is
\begin{equation}
\nonumber
\sum_{j\in N} \left( d_j(b_1-c) + \sum_{\substack{w \in N \\l(j,w)>1}}{b_{l(j,w)}} \right)
 \end{equation}

\begin{lemma}
\label{lem:efficient}
Let $\mu$ be the number of nodes in the network. \\
(i) If $c < b_1-b_2$, then complete graph is uniquely efficient.\\
(ii) If $b_1-b_2< c \leq b_1 + \left( \frac{\mu-2}{2} \right) b_2$, then star is the unique efficient topology.\\
(iii) If $c >b_1+ \left( \frac{\mu-2}{2}\right)b_2 $, then null graph is uniquely efficient.\\
\end{lemma}
\noindent
Lemma~\ref{lem:efficient} follows from the analysis of efficient networks by Narayanam and Narahari~\cite{ramasuri1}.

The null graph in the proposed model of recursive network formation corresponds to a single node to which no other node prefers to connect, and so the network does not grow. 

\begin{theorem}
\label{thm:eff_star}
Based on the derived sufficient conditions, null graph, star graph, and complete graph are efficient.
\end{theorem}
\begin{proofblackdot}
It is easy to see that irrespective of the value of $c_0$, if $c>b_1$, no node, external to the network, connects to the only node in the network and hence, does not enter the network. Such a network is trivially efficient as in the range $c>b_1$, it is a star of one node and also a null graph. It is also clear that the star topology and the complete graph are efficient as the conditions on $c$ from Theorems~\ref{thm:star} and \ref{thm:complete} form a subset of the range of $c$ in which these topologies are respectively efficient.
\end{proofblackdot}

Theorems~\ref{thm:eff_bipartite} and \ref{thm:eff_kstar} give bounds on the efficiency of bipartite Tur\'an network and $k$-star network, respectively. We provide their proofs in the Appendix.

\begin{theorem}
\label{thm:eff_bipartite}
Based on the derived sufficient conditions, for $\mu$ sufficiently large, efficiency of bipartite Tur\'an network is half of that of the efficient network in the worst case and the network is close to being efficient in the best case.
\end{theorem}

\begin{proofblackdot} 
Here, we make a reasonable assumption that the number of nodes in the network is large. Hence we can assume that $\mu$ is even without loss of accuracy. The sum of utilities of nodes in a bipartite Tur\'an network with even number of nodes is
\begin{equation}
\nonumber
2\left( \frac{\mu}{2} \right)^2 (b_1-c)+2 \dbinom{\frac{\mu}{2}}{2}2b_2
\end{equation}
From Lemma~\ref{lem:efficient}, star network is efficient in the range of $c$ derived in Theorem~\ref{thm:bipartite}. So, to get the efficiency of the bipartite Tur\'an network relative to the star network, we divide the above sum by the sum of utilities of nodes in a star network, which is
\begin{equation}
\label{eqn:star_eff}
2(\mu-1)(b_1-c)+\dbinom{\mu -1}{2}2b_2
\end{equation}
Using the assumption that $\mu$ is large and the fact from the sufficient conditions that $b_2$ is comparable to $b_1-c$, it can be shown that the efficiency relative to the star network, approximately is
\begin{equation}
\nonumber
\frac{1}{2}+\frac{b_1-c}{2b_2}
\end{equation}
As the range of $c$ in Theorem~\ref{thm:bipartite} depends on the value of $\gamma$, the values of $c$ are bounded by $b_1-b_2$ and $b_1-b_3$. So the efficiency is bounded by 1 and $\left( \frac{1}{2}+\frac{b_3}{2b_2} \right)$ of that of the star network, which can take a minimum value of $\frac{1}{2}$ of that of star network when $b_3<<b_2$. 
\end{proofblackdot}

\begin{theorem}
\label{thm:eff_kstar}
Based on the derived sufficient conditions, for $\mu$ sufficiently large, efficiency of $k$-star network is $\frac{1}{k}$ of that of the efficient network in the worst case and the network is close to being efficient in the best case.
\end{theorem}

\begin{proofblackdot}
We make the reasonable assumption of $\mu$ being large. In particular, $\mu >> k$ (not necessarily $>>k^2$).
Hence we can assume that $\mu$ is divisible by $k$ without loss of accuracy. The sum of utilities of nodes in such a $k$-star network is
\begin{equation}
\nonumber
\begin{split}
2\dbinom{k}{2}(b_1-c)+2(\mu-k)(b_1-c)+ 2k \left( 1-\frac{1}{k} \right) (\mu -k) b_2\\+2k \dbinom{\frac{\mu}{k} -1}{2}b_2 + k \left( \frac{\mu}{k}-1 \right) \left( 1- \frac{1}{k} \right) (n-k) b_3
\end{split}
\end{equation}
From Lemma~\ref{lem:efficient}, star network is efficient in the range of $c$ derived in Theorems~\ref{thm:2star} and \ref{thm:kstar}. So, to get the efficiency of the $k$-star network relative to the star network, we divide the above sum by Expression~(\ref{eqn:star_eff}).
Using the assumption that $\mu$ is large and the fact from the sufficient conditions that $b_2$ and $b_3$ are comparable to $b_1-c$, it can be shown that the efficiency relative to the star network, approximately is
\begin{equation}
\nonumber
\frac{1}{k}+ \left( 1- \frac{1}{k} \right) \frac{b_3}{b_2}
\end{equation}
As $b_3$ is bounded by $0$ and $b_2$, the efficiency of $k$-star is bounded by $\frac{1}{k}$ and 1 of that of the star network.
\end{proofblackdot}

\section{Discussion and Future work}
\label{sec:conclusion}

We proposed a model of recursive network formation where nodes enter a network sequentially, thus triggering evolution of the network each time a new node enters.
Though we have assumed a sequential move game model with myopic nodes and pairwise stability as the solution concept, the model, as depicted in Figure~\ref{fig:model}, is independent of the model of network evolution, the solution concept used for equilibrium state, and also the utility model.
The recursive nature of our model enabled us to directly analyze the network formation game using
an elegant induction based technique.
We derived sufficient conditions by directing the dynamics of network formation along a desired improving path in the sequential move game tree.

\subsection{Future work}
\label{sec:future}
This work proposed a way to derive conditions on cost for maintaining link with an immediate neighbor and also for entering a network, under which a desired network topology is obtained. Going a step further, it would be interesting to design incentives such that agents in a network comply with these conditions. 
The proposed model of network formation can be extended to other utility models to investigate the formation of interesting topologies under them.
Our analysis ensures that irrespective of the chosen node at any point in time, the network evolution is directed as desired.
A possible solution for simplifying the analysis for more involved topologies is to carry out probabilistic analysis for deriving conditions so that a network has the desired topology with high probability.
Another interesting direction, from a practical viewpoint, is to study the problem of forming networks where the topology need not be exactly the one which is ideally desirable, for example, a near-$k$-star network instead of a precise $k$-star.

\section*{Acknowledgments}

The original publication to appear in the Proceedings of The 8th Workshop on Internet \& Network Economics, titled \textit{Forming Networks of Strategic Agents with Desired Topologies}, will be soon available at \href{www.springerlink.com}{www.springerlink.com}.
An extended version of this paper is under review in IEEE Transactions on Network Science and Engineering.
The authors thank Rohith D. Vallam for useful suggestions.

\end{spacing}

\begin{spacing}{.9}

\bibliographystyle{abbrv}
\bibliography{Sufficient_Conditions_for_Formation_of_a_Network_Topology_by_Self-interested_Agents_references}  
\end{spacing}
\begin{spacing}{1}

\newpage
\appendix
\section*{APPENDIX}

\noindent
\textbf{Theorem~\ref{thm:bipartite}.}
\textit{
For a network with $\gamma <   \frac{b_2 - b_3}{3b_2 - b_3} $, if $b_1-b_2+ \gamma \left( 3b_2 - b_3 \right) <  c < b_1 - b_3$ 
and $\left( 1-\gamma \right) \left( b_2-b_3 \right) < c_0 \leq \left( 1-\gamma \right) b_2$, the resulting topology is a 
bipartite Tur\'an graph.
}\\

\begin{proofblackdot}
We first derive conditions for pairwise stability of a bipartite Tur\'an network, that is assuming that such a network is formed, what conditions are required so that there are no incentives for any two unconnected nodes to create a link between them and for any node to delete any of its links. Note that these conditions can be integrated in the later part of the proof within different scenarios that we consider.\\
In what follows, $p_1$ is the size of the partition constituting the node taking its decision, $p_2$ is the size of the other partition and $n=p_1+p_2$ is the number of nodes in the network.
We need to consider cases for some discretely small number of nodes owing to the nature of essential nodes, after which the analysis holds for arbitrarily large number of nodes. For brevity, we present the analysis for the base case and a generic case in each scenario, omitting presentation of discrete cases. \\

\noindent
\textbf{No two nodes belonging to the same partition should create a link between them:} That is, their utility should not increase by doing so. This is not applicable for $n=2$. \\
For $n=3$, 
\begin{equation}
\nonumber
2(b_1-c) \leq b_1-c+(1-\gamma)b_2
\end{equation}
\begin{equation}
\label{E16}
\Leftrightarrow c \geq b_1-b_2+\gamma b_2
\end{equation}
For $n\geq 4$,
\begin{equation}
\nonumber
(p_2+1)(b_1-c)+(p_1-2)b_2 \leq p_2(b_1-c)+(p_1-1)b_2
\end{equation}
\begin{equation}
\nonumber
\Leftrightarrow c \geq b_1-b_2
\end{equation}
which is a weaker condition that Inequality~(\ref{E16}).\\

\noindent
\textbf{No node should delete its link with any node belonging to the other partition:} That is, their utility should not increase by doing so. \\
For $n=2$,
\begin{equation}
\nonumber
0\leq b_1-c
\end{equation}
\begin{equation}
\label{E17for2}
\Leftrightarrow c \leq b_1
\end{equation}
For $n\geq 6$,
\begin{equation}
\nonumber
(p_2-1)(b_1-c)+(p_1-1)b_2+b_3 \leq p_2(b_1-c)+(p_1-1)b_2
\end{equation}
\begin{equation}
\label{E17}
\Leftrightarrow c \leq b_1-b_3
\end{equation}
It can be shown that conditions for the discrete cases $n=3,4,5$  are satisfied by Inequality~(\ref{E17}).\\  \\
In the process of formation of a bipartite Tur\'an network, at most four different types of nodes exist at any point in time.
  \begin{center}
\begin{tabular}{| l | l |}
    \hline
\T \B  I & newly entered node \\ \hline
\T \B  II & nodes connected to the newly entered node\\ \hline
\T \B  III & nodes in the same partition as II, but not of Type II\\ \hline
\T \B  IV & rest of the nodes\\ \hline
  \end{tabular}
  \end{center}
The notation we use while deriving the sufficient conditions are as follows:
\begin{center}
  \begin{tabular}{| l | l |}
    \hline
\T \B $k$ & number of nodes of Type II \\ \hline
\T \B $n$ & number of nodes in the network, including new node\\ \hline
\T \B $m_1$ & number of nodes of Types II and III put together\\ \hline
\T \B $m_2$ & number of nodes of Type IV\\ \hline
  \end{tabular}
\end{center}

\noindent
\textbf{For the newly entering node to enter the network:} Its utility should be positive after doing so. Also, in case of even $n$, for the new node to be a part of the smaller partition, its first connection should be a node belonging to the larger partition. So for $k=0$, we have \\
For $n\geq 2$,
\begin{equation}
\nonumber
b_1-c+ \lceil \frac{n}{2}-1 \rceil \left( (1-\gamma)b_2 - c_0 \right) + \lfloor \frac{n}{2}-1 \rfloor (1-\gamma)b_3 >0
\end{equation}
It can be seen that the condition is the strongest when $n=2$ whenever
\begin{equation}
\label{E1b}
c_0 \leq (1-\gamma)b_2
\end{equation}
The condition thus becomes
\begin{equation}
\nonumber
c< b_1
\end{equation}
which is satisfied by Inequality~(\ref{E17}).\\

\noindent 
\textbf{The utility of a node in the larger partition, whenever applicable, should not decrease after accepting link from the new node:}\\
For $n=2$,
\begin{equation}
\nonumber
b_1-c \geq 0
\end{equation}
\begin{equation}
\nonumber
\Leftrightarrow c \leq b_1
\end{equation}
For $n \geq 5$,
\begin{equation}
\nonumber
\lceil \frac{n}{2} \rceil (b_1 -c) + \lfloor \frac{n}{2}-1 \rfloor b_2 + \gamma \lceil \frac{n}{2}-1 \rceil 2 b_2 + \gamma \lfloor \frac{n}{2}-1 \rfloor 2 b_3 
\geq \lceil \frac{n}{2}-1 \rceil (b_1 -c) + \lfloor \frac{n}{2}-1 \rfloor b_2
\end{equation}
\begin{equation}
\nonumber
\Leftrightarrow c \leq b_1+ 2\gamma \lceil \frac{n}{2}-1 \rceil b_2 + \lfloor \frac{n}{2}-1 \rfloor b_3
\end{equation}
The conditions for these as well as the discrete cases $n=3,4$ are satisfied by Inequality~(\ref{E17}).\\

\noindent
\textbf{The new node should connect to a node in the larger partition, whenever applicable:} One way to see this is by ensuring that this strategy strictly dominates connecting to a node in the smaller partition.
This scenario arises for even values of $n\geq 4$.
\begin{equation}
\begin{split}
\nonumber
b_1-c+ \left( \frac{n}{2}-1 \right) \left( (1-\gamma)b_2 - c_0 \right) + \left( \frac{n}{2}-1 \right) (1-\gamma)b_3 \\
> b_1-c+ \left( \frac{n}{2} \right) \left( (1-\gamma)b_2 - c_0 \right) + \left( \frac{n}{2}-2 \right) (1-\gamma)b_3 
\end{split}
\end{equation}
\begin{equation}
\label{E2a}
\Leftrightarrow c_0 > (1-\gamma)(b_2-b_3)
\end{equation}
An alternative condition would be such that the utility of a node in the smaller partition decreases if it accepts the link from the new node, thus forcing the latter to connect to a node in the other partition. But it can be seen that this condition is inconsistent with Inequality~(\ref{E17}) and so we use Inequality~(\ref{E2a}) to meet our purpose.\\

\noindent
\textbf{Type I node should prefer connecting to a Type III node, if any, than remaining in its current state:}
For $k\geq 2$, this scenario does not arise for $n<6$. 
For $n\geq 6$,
\begin{equation}
\nonumber
(k+1)(b_1-c)+m_2b_2+(m_1-k-1)b_3 > k(b_1-c)+m_2b_2+(m_1-k)b_3
\end{equation}
\begin{equation}
\label{E6}
\Leftrightarrow c<b_1-b_3
\end{equation}
Now for $k=1$, this scenario does not arise for $n=2,3$.\\
For $n \geq 4$,
\begin{equation}
\nonumber
2(b_1-c)+m_2b_2+(m_1-2)b_3 > b_1-c+(1-\gamma)m_2b_2+(1-\gamma)(m_1-1)b_3
\end{equation}
\begin{equation}
\nonumber
\Leftrightarrow c<b_1-b_3+\gamma(m_2b_2+(m_1-1)b_3)
\end{equation}
Note that as $n \geq 4$, we have $m_1 \geq 2$ and $m_2 \geq 1$ and so the above condition is weaker that Inequality~(\ref{E6}).\\
It is also necessary that utility of Type III node does not decrease on accepting link from Type I node. In fact, when the former gets a chance to move, we derive conditions so that it also volunteers to create a link with the later.\\

\noindent
\textbf{The utility of Type III node should increase if it successfully creates a link with Type I node:}
When $k=1$, the case does not arise for $n=2,3$.\\
For $n \geq 6$,
\begin{equation}
\nonumber
(m_2+1)(b_1-c)+(m_1-1)b_2>m_2(b_1-c)+(m_1-1)b_2+(1-\gamma)b_3
\end{equation}
\begin{equation}
\nonumber
\Leftrightarrow c<b_1-b_3+\gamma b_3
\end{equation}
The conditions obtained from discrete cases $n=4,5$ are weaker than this one.\\
For $k\geq 2$, this case does not arise for $n <6$. \\
For $n\geq 6$,
\begin{equation}
\nonumber
(m_2+1)(b_1-c)+(m_1-1)b_2>m_2(b_1-c)+(m_1-1)b_2+b_3
\end{equation}
\begin{equation}
\nonumber
\Leftrightarrow c<b_1-b_3
\end{equation}
The conditions for all cases are satisfied by Inequality~(\ref{E6}).\\

\noindent
\textbf{Type III node should not delete its link with Type IV node:} This can be assured if this strategy is dominated by its strategy of forming a link with Type I node.
This scenario does not arise for $n=2,3$. 
The conditions for the discrete cases $n=4,5,6$ are weaker than that for $n\geq 7$.\\
For $n\geq 7$,
\begin{equation}
\nonumber
(m_2+1)(b_1-c)+(m_1-1)b_2>(m_2-1)(b_1-c)+b_3+(m_1-1)b_2+(1-\gamma)b_3
\end{equation}
\begin{equation}
\nonumber
\Leftrightarrow c<b_1-b_3+\frac{\gamma}{2}b_3
\end{equation}
For $k\geq 2$, the cases applicable are $n\geq 6$.
The condition for discrete case $n=6$ is weaker than the following condition.\\
For $n\geq 7$,
\begin{equation}
\nonumber
(m_2+1)(b_1-c)+(m_1-1)b_2>(m_2-1)(b_1-c)+b_3+b_3+(m_1-1)b_2
\end{equation}
\begin{equation}
\nonumber
\Leftrightarrow c<b_1-b_3
\end{equation}
Hence, all conditions for this scenario are satisfied by Inequality~(\ref{E6}).\\

\noindent
\textbf{Type III node should prefer connecting to Type I node than to another Type III node:} This does not arise for $n<6$.
When $k=1$, \\
For $n\geq 6$,
\begin{equation}
\nonumber
(m_2+1)(b_1-c)+(m_1-1)b_2>(m_2+1)(b_1-c)+(m_1-2)b_2+(1-\gamma)b_3
\end{equation}
\begin{equation}
\nonumber
\Leftrightarrow b_2>(1-\gamma)b_3
\end{equation} 
which is always true. For $k\geq 2$,\\
For $n \geq 6$,
\begin{equation}
\nonumber
(m_2+1)(b_1-c)+(m_1-1)b_2>(m_2+1)(b_1-c)+(m_1-2)b_2+b_3
\end{equation}
\begin{equation}
\nonumber
\Leftrightarrow b_2>b_3
\end{equation} 
which is always true.\\

\noindent
\textbf{Type IV node should not delete its link with Type III node:} That is, its utility should not increase by doing so.
This does not arise for $n<4$.\\
For $n\geq 7$,
\begin{equation}
\nonumber
(m_1-1)(b_1-c)+(m_2-1)b_2+(1-\gamma)b_2+b_3 \leq m_1(b_1-c)+(m_2-1)b_2+(1-\gamma)b_2
\end{equation}
\begin{equation}
\nonumber
\Leftrightarrow  c\leq b_1-b_3
\end{equation} 
The conditions for discrete cases $n=4,5,6$ are weaker than the above condition.
For $k\geq 2$, the new cases are $n\geq 6$, where the discrete case $n=6$ result in conditions weaker than the following one.\\
For $n \geq 7$,
\begin{equation}
\nonumber
(m_1-1)(b_1-c)+(m_2-1)b_2+b_2+b_3 \leq m_1(b_1-c)+(m_2-1)b_2+b_2
\end{equation}
\begin{equation}
\nonumber
\Leftrightarrow c\leq b_1-b_3
\end{equation} 
It can be seen that all conditions of this scenario are satisfied by Inequality~(\ref{E6}).\\

\noindent
\textbf{Type IV node should also not break its link with Type II node:} That is, its utility should not increase by doing so.
For $k=1$,\\
For $n\geq 6$,
\begin{equation}
\nonumber
(m_1-1)(b_1-c)+(m_2-1)b_2+(1-\gamma)b_4+(1-\gamma)b_3  \leq m_1(b_1-c)+(m_2-1)b_2+(1-\gamma)b_2 
\end{equation}
\begin{equation}
\nonumber
\Leftrightarrow c\leq b_1-b_3+(1-\gamma)(b_2-b_4)+\gamma b_3
\end{equation} 
The discrete cases $n=3,4,5$ result in weaker conditions than this.
For $k\geq 2$,\\
For $n\geq 6$,
\begin{equation}
\nonumber
(m_1-1)(b_1-c)+(m_2-1)b_2+b_2+b_3  \leq m_1(b_1-c)+(m_2-1)b_2+b_2 
\end{equation}
\begin{equation}
\nonumber
\Leftrightarrow c\leq b_1-b_3
\end{equation} 
The conditions are satisfied by Inequality~(\ref{E6}).\\

\noindent
\textbf{Type I node should not propose a link to a Type IV node:} One way is to ensure that this strategy of Type I node is dominated by its strategy to propose a link to a Type III node.
It can be seen that for $k \geq 2$ and $n \geq 6$, this translates to
\begin{equation}
\nonumber
(k+1)(b_1-c)+m_2b_2+(m_1-k-1)b_3> (k+1)(b_1-c)+(m_2-1)b_2+(m_1-k)b_2
\end{equation}
\begin{equation}
\nonumber
\Leftrightarrow b_2-b_3>(m_1-k)(b_2-b_3)
\end{equation}
which is not true for $m_1>k$.\\
So we look at the alternative condition that the utility of Type IV node decreases if it accepts the link from Type I node, and as Type I node computes this decrease in utility, it will not propose a link to Type IV node.
First, we consider $k=1$. The discrete case $n=4$ gives the following condition.\\
\begin{equation}
\nonumber
3(b_1-c)+2\gamma b_2+2\gamma b_2 < 2(b_1-c)+(1-\gamma)b_2+2\gamma b_2+\gamma b_3 
\end{equation}
\begin{equation}
\label{E7b}
\Leftrightarrow c>b_1-b_2+\gamma(3b_2-b_3)
\end{equation}
The other discrete cases $n=3,5$ result in weaker conditions than the above.\\
For $n\geq 6$,
\begin{equation}
\nonumber
(m_1+1)(b_1-c)+(m_2-1)b_2<m_1(b_1-c)+(m_2-1)b_2+(1-\gamma)b_2
\end{equation}
\begin{equation}
\nonumber
\Leftrightarrow c>b_1-b_2+\gamma b_2 
\end{equation}
which is a weaker condition than Inequality~(\ref{E7b}).
Now for $k \geq 2$, $n=4,5$ correspond to pairwise stability conditions and cases $n<4$ are not applicable.\\
For $n\geq 6$,
\begin{equation}
\nonumber
(m_1+1)(b_1-c)+(m_2-1)b_2 < m_1(b_1-c)+(m_2-1)b_2+b_2
\end{equation}
\begin{equation}
\nonumber
\Leftrightarrow c>b_1-b_2
\end{equation}
which is satisfied by Inequality~(\ref{E7b}).\\

\noindent
\textbf{Type IV node should not propose a link to Type I node:} This scenario is essentially equivalent to the previous one scenario of utility of Type IV node decreasing due to link with Type I node, with the equalities permitted. So these result in weaker and hence no additional conditions.\\

\noindent
\textbf{Type III node should not propose a link to Type II node:} One way is to ensure that for Type III node, connecting to Type II node is strictly dominated by connecting to Type I node.
It can be seen that for $k\geq 2$ and $n\geq 6$, this translates to
\begin{equation}
\nonumber
(m_2+1)(b_1-c)+(m_1-2)b_2+b_2<(m_2+1)(b_1-c)+(m_1-1)b_2
\end{equation}
which gives $0>0$. 
So we need to use the alternative condition that the utility of Type II node decreases on accepting link from Type III node.
For $k=1$,\\
For $n=4$,
\begin{equation}
\nonumber
3(b_1-c)+4\gamma b_2 < 2(b_1-c)+(1-\gamma)b_2+2\gamma b_2+\gamma b_3
\end{equation}
\begin{equation}
\nonumber
\Leftrightarrow c > b_1-b_2+\gamma(3b_2 - b_3)
\end{equation}
which is same as Inequality~(\ref{E7b}). \\
For $n\geq 5$,
\begin{equation}
\nonumber
\begin{split}
(m_2+2)(b_1-c)+(m_1-2)b_2+2\gamma(m_2+1)b_2+2\gamma(m_1-2)b_3 \\
< (m_2+1)(b_1-c)+(m_1-1)b_2+2\gamma m_2b_2+ 2\gamma(m_1-1)b_3
\end{split}
\end{equation}
\begin{equation}
\nonumber
\Leftrightarrow c > b_1-b_2+2\gamma(b_2-b_3)
\end{equation}
which is a weaker condition than Inequality~(\ref{E7b}). 
Now for $k \geq 2$, the only new case is the following.\\
For $n\geq 6$,
\begin{equation}
\nonumber
(m_2+2)(b_1-c)+(m_1-2)b_2<(m_2+1)(b_1-c)+(m_1-1)b_2
\end{equation}
\begin{equation}
\nonumber
\Leftrightarrow c>b_1-b_2
\end{equation}
which is satisfied by Inequality~(\ref{E7b}).\\

\noindent
\textbf{Type II node should not propose a link with Type III node:} This is essentially equivalent to the above scenario of utility of Type II node decreasing due to link with Type III node, with the equalities permitted. So these result in weaker and hence no additional conditions.\\

\noindent
\textbf{No Type II node should delete link with Type IV node:} First, we consider $k=1$.\\
For $n \geq 7$, 
\begin{equation}
\nonumber
\begin{split}
(m_2-1)(b_1-c) + (b_1-c) + (m_1-1)b_2+b_3+2\gamma (m_2-1)b_2+2\gamma (m_1-1)b_3 +2\gamma b_4 \\ \leq m_2(b_1-c)+(b_1-c) + (m_1-1)b_2+2\gamma m_2 b_2 + 2\gamma(m_1-1)b_3
\end{split}
\end{equation}
\begin{equation}
\nonumber
\Leftrightarrow c \leq b_1 - b_3 + 2 \gamma (b_2 - b_4)
\end{equation}
This as well as all discrete cases $n<7$ are satisfied by Inequality~(\ref{E6}).\\
For $k \geq 2$,  the cases of $n=4,5$ correspond to pairwise stability condition that we have already considered, while cases $n<4$ are not applicable.\\
For $n \geq 6$,
\begin{equation}
\nonumber
m_2(b_1-c)+(m_1-1)b_2+b_3 \leq (m_2+1)(b_1-c)+(m_1-1)b_2
\end{equation}
\begin{equation}
\nonumber
\Leftrightarrow c \leq b_1-b_3
\end{equation}
which is satisfied by Inequality~(\ref{E6}).\\

\noindent
\textbf{Two Type IV nodes should not create a mutual link:} That is their utilities should not increase by doing so.
When $k=1$, it is not applicable for $n<5$. Also, the discrete case $n=5$ results in the same condition as below.\\
For $n\geq 6$,
\begin{equation}
\nonumber
(m_1+1)(b_1-c)+(m_2-2)b_2+(1-\gamma)b_2 \leq m_1(b_1-c)+(m_2-1)b_2+(1-\gamma)b_2
\end{equation}
\begin{equation}
\nonumber
\Leftrightarrow c \geq b_1-b_2
\end{equation} 
For $k\geq 2$, $n=5$ corresponds to pairwise stability condition. \\
For $n\geq 6$,
\begin{equation}
\nonumber
(m_1+1)(b_1-c)+(m_2-2)b_2+b_2 \leq m_1(b_1-c)+(m_2-1)b_2+b_2
\end{equation}
\begin{equation}
\nonumber
\Leftrightarrow c \geq b_1-b_2
\end{equation} 
These are weaker conditions than Inequality~(\ref{E7b}).\\

\noindent
\textbf{No two Type II nodes should create a link between themselves:} This only applies to $k\geq 2$.
Also $n=4,5$ result in pairwise stability condition.\\
For $n\geq 6$,
\begin{equation}
\nonumber
(m_2+2)(b_1-c)+(m_1-2)b_2 \leq (m_2+1)(b_1-c)+(m_1-1)b_2
\end{equation}
\begin{equation}
\nonumber
\Leftrightarrow c\geq b_1-b_2
\end{equation} 
which is a weaker condition than Inequality~(\ref{E7b}).\\

\noindent
\textbf{Link between Type I node and Type II node  should not be deleted:} It is clear that it will not be deleted as such a link is just formed with no other changes in the network.\\

Inequalities~(\ref{E6}) and (\ref{E7b}) are stronger conditions than Inequalities~(\ref{E16}), (\ref{E17for2}) and (\ref{E17}).
Furthermore, for non-zero range of $c$, from Inequalities~(\ref{E6}) and (\ref{E7b}), we have
\begin{equation}
\label{E0}
\gamma < \frac{b_2-b_3}{3b_2-b_3}
\end{equation} 
The required sufficient conditions are obtained by combining Inequalities~(\ref{E1b}), (\ref{E2a}), (\ref{E6}), (\ref{E7b}) and (\ref{E0}).
\end{proofblackdot}

\end{spacing}

\end{document}